\documentclass[preprint,12pt,3p]{singlecol-new}
\usepackage{natbib,stfloats}
\usepackage{mathrsfs}
\usepackage{amsmath,amssymb,amsfonts}
\usepackage{algorithmic}
\usepackage{graphicx}
\usepackage{scrtime}
\usepackage{float}
\usepackage[dont-mess-around]{fnpct}
\usepackage{etoolbox}
\newcommand*{\affmark}[1][*]{\textsuperscript{#1}}
\usepackage{textcomp}
\usepackage{xcolor}
\def\BibTeX{{\rm B\kern-.05em{\sc i\kern-.025em b}\kern-.08em
		T\kern-.1667em\lower.7ex\hbox{E}\kern-.125emX}}

\theoremstyle{TH}{

}

\theoremstyle{THrm}{

}

\theoremstyle{THhit}{

}

\makeatletter
\def\theequation{\arabic{equation}}

\makeatother

\begin{document}%
%%%%%%%%%%%%%%%%%

\setcounter{page}{1}

\LRH{S. M. U. Hashmi  et~al.}
%\LRH{Author  et~al.}

\RRH{Energy Efficient Cross Layer Time Synchronization in Cognitive Radio Networks}

\VOL{x}

\ISSUE{x}

\PUBYEAR{201X}

\BottomCatch

\CLline

\subtitle{}

\title{Energy Efficient Cross Layer Time Synchronization in Cognitive Radio Networks}

\authorA{S.M. Usman Hashmi \protect\affmark[1], Muntazir Hussain \protect\affmark[1],   S.M. Nashit Arshad \protect\affmark[1]}
\affA{\affmark[1] Department of Electronics Engineering\\ Iqra University, Islamabad, Pakistan \\
	E-mail: usman16288@yahoo.com, \\
	E-mail: muntazir\_hussain14@yahoo.com,  \\
	E-mail: nashit@iqraisb.edu.pk}
\authorB{Kashif Inayat\protect\affmark[2]}
\affB{\affmark[2] Department of Electronics  Engineering,\\ Incheon National University, South Korea \\
	E-mail: kashif.inayat@inu.ac.kr}
%

%
%\authorA{Fusheng Wang\footnote{Work done while working at Siemens Corporate Research.} }
%
%\affA{Department of Biomedical Informatics, Emory University
%\newline
%36 Eagle Row, Ste 589, Atlanta, GA 30322, USA}
%
%
%
%\authorB{\footnotesize Cristobal Vergara-Niedermayr\footnote{Work done while working at Siemens Corporate Research.}}
%\affB{Oracle \newline
%New Jersey, USA}
%
%
\authorC{Seong Oun Hwang\protect\affmark[3,*]}
\affC{\affmark[3] Department of Computer Engineering,
	\\ Gachon University, South Korea\\
	Email: sohwang@gachon.ac.kr\\\affmark[*]{Corresponding author} }

%
%
%\authorB{Kashif Inayat\protect\affmark[2]}
%\affB{\affmark[2] Department of Electronics and Computer Engineering,\\ Hongik University, Sejong, Korea \\
%E-mail: kashif\_chaudhary@yahoo.com}
%\authorC{Seong Oun Hwang\protect\affmark[3,*]}
%\affC{\affmark[3] Department of Computer Engineering,\\ Gachon University, Korea\\
%Email: sohwang@gachon.ac.kr\\\affmark[*]{Corresponding author: 	Seong Oun Hwang,  sohwang@gachon.ac.kr} }
%%{Corresponding author: Seong Oun Hwang (sohwang@gachon.ac.kr)}

\begin{abstract}
Time synchronization is a vital concern for any Cognitive Radio Network (CRN) to perform dynamic spectrum management. Each Cognitive Radio (CR) node has to be environment aware and self adaptive and must have the ability to switch between multiple modulation schemes and frequencies. Achieving same notion of time within these CR nodes is essential to fulfill the requirements for simultaneous quiet periods for spectrum sensing. Current application layer time synchronization protocols require multiple timestamp exchanges to estimate skew between the clocks of CRN nodes. The proposed symbol timing recovery method already estimates the skew of hardware clock at the physical layer and use it for skew correction of application layer clock of each node. The heart of application layer clock is the hardware clock and hence application layer clock skew will be same as of physical layer and can be corrected from symbol timing recovery process. So one timestamp is enough to synchronize two CRN nodes. This conserves the energy utilized by application layer protocol and makes a CRN energy efficient and can achieve time synchronization in short span.
\end{abstract}

\KEYWORD{time synchronization, cognitive radio, clock correction, symbol timing recovery}

\REF{to this paper should be made as follows: Hashmi, S. M. U., Hussain, M., Arshad, S. M . N., Inayat, K., and Hwang, S. O. (2019)  `Energy Efficient Cross Layer Time Synchronization in Cognitive Radio Networks', {\it Int. J. Internet Technology and Secured Transactions}, Vol. x, No. x, pp.xxx\textendash xxx.}

\begin{bio}
S.M. Usman Hashmi received his PhD degree in the domain of Telecommunication and Networks. He is an Assistant Professor in, Electronics Engineering Department at Iqra University, Islamabad, Pakistan. His research interest includes signal processing, communication, and time synchronization at the physical layer.  \vs{9}

\noindent Muntazir Hussain received his PhD degree in the domain of Electrical, Engineering. He  is an Assistant Professor in Electronics Engineering, Department at Iqra University, Islamabad, Pakistan. His research interest, includes the controller and anti-windup compensator design, time-delay system, non-linear control system, and stability of the power control system. \vs{8}

\noindent S.M. Nashit Arshad received his Bachelor degree of Mechatronics from Air, University with major in Robotics, his Master in Mechatronics with specialization in control systems from Air University. He served various Industries in the, capacity of control engineer, design engineer and production engineer. Nowadays, he is serving as a lecture in electronics department and he has also started his Ph.D in robotics with specialization in artificial intelligence from NUST Islamabad.\vs{7}

\noindent Kashif Inayat received his BE degree in electronics engineering from Iqra Univeristy Islamabad Campus, Pakistan in 2014, his MS degree in electronics and computer Engineering in 2019 from Hongik University, South Korea. He worked as a engineer at Digital System Design Lab at Iqra University Islamabad, from 2014 to 2017. He also worked as a researcher at R\&D Institute Incheon National University in collaboration with Samsung Electronics, South Korea. He is currently pursuing his PhD in electronics engineering at Incheon National University. His research agenda is centered on neuromorphic, embedded systems and information security. He borrow doctrine from information security, learn from machine learning and apply electronics concepts to drive his research in the following applications: AI-applications specific integrated circuits (ASIC's), tensor processing, information security, and blockchain.\vs{6}

\noindent Seong Oun Hwang received his BS degree in Mathematics in 1993 from Seoul National University, his MS degree in Information and Communications Engineering in 1998 from Pohang University of Science and Technology, and his PhD degree in Computer Science from Korea Advanced Institute of Science and Technology, Republic of Korea.He worked as a software engineer at LG-CNS Systems, Inc. from 1994 to 1996. He worked asa senior researcher at Electronics and Telecommunications Research Institute (ETRI) from 1998 to 2007. He worked as a professor with the Department of Software and Communications Engineering at Hongik University from 2008 to 2019. He is currently a professor of the Department of Computer Engineering at Gachon University and an editor of ETRI Journal. His research interests include cryptography, cyber security and artificial intelligence.\\
This paper is a revised and expanded version of a paper entitled, "Optimized Time Synchronization for Cognitive Radio Networks",
presented at  {\it International conference on green and human information technology 2019},  Kuala Lumpur, Malaysia, Jan 16-18, 2019.

\end{bio}

\maketitle

\section{Introduction}
Cognitive radio (CR) is a promising and keen wireless communication system that is environment aware and can adapt its functioning parameters in real time. These parameters may include transmit power, modulating frequency and modulation scheme (\cite{b1,b2,b3,b4,b5}). Parameter adaptation results in dynamic spectrum access which allow an unlicensed user to temporary utilize the unused licensed user frequency band. Dynamic spectrum allocation increases the overall spectral efficiency and capacity (\cite{b6,b7,b8}). Spectrum sensing, primary user interference management and allocation of resources must be efficiently handled (\cite{b9, b10, b11}). The implementation of cognitive radio requires Software Defined Radio (SDR) hardware on which the parameter of radio communication can vary its parameters in real time (\cite{b12,b3,b4}).
	
Current communication networks do not have the ability to adapt to network changes which results in inefficient use of spectrum. It has fixed protocols and policies and is unable to make any intelligent adaptation. Communication network that is based on cognitive radio technique and can sense the network parameters and adapt accordingly is call Cognitive Radio Network (CRN) (\cite{b9,b12,b7,b10}). These networks have higher spectral efficiency and more number of users can accommodate in the network. Challenges in implementation of CRN include spectrum sensing, spectrum management, use of unlicensed spectrum, spectrum sharing techniques, hidden node problem, security, adaptable radio interface, hardware and software architecture and time synchronization. Spectrum sensing has to be done in such a way that it should not interfere with primary user data. The modulation type, power or frequency are the useful parameters for the detection of available slots in the spectrum. Available techniques are based on energy or feature detection. Spectrum management must be efficient and can adapt to different transmission power and frequency. Spectrum sharing is allocation of spectrum to unlicensed user for shared services. Hidden node problem must be addressed in order to avoid interference with primary user. Security of data communicated is very essential because data has to be sent over primary user's spectrum and that user might be able to access that data. The radio interface should be adaptable to network changes and hardware and software architectures must be designed accordingly. 

The motivation behind this works is that one of the major challenges in any CRN is time synchronization with in the nodes of the network (\cite{b13,b14,b15}). Each node must be aware of spectrum changes at every instance and reliable sensing is only possible in a time synchronized network. A promising protocol that must be energy efficient, needs to be developed which can synchronize two nodes with lesser number of timestamp exchanges.

CRN can be implemented in many current infrastructures e.g., cellular networks and smart grid. Smart Grid Network is a three layered network. The physical layer deals with the generation and distribution of power. Network layer is responsible for communication in the network. The applications layer provides the services such as advanced metering, demand response and grid management. Bandwidth requirements of this network are estimated to be in 10kbps to 100 kbps range per device. Bandwidth becomes a challenge as the data traffic grows dramatically. Cognitive radio based smart grid may overcome the bandwidth issue by efficient utilization of spectrum. Public Safety Networks are getting popular to provide emergency services. It can prevent or helps responding to a particular incident. This network requires wireless laptops, hand held devices, and video surveillance cameras to have wireless services such as messaging, email, picture transfer, video streaming etc. This will eventually increase bandwidth requirement and cognitive radio can increase bandwidth efficiency. Cellular networks is always in need of bandwidth because of increase in number of users and services provided. Cognitive radio technology can be used with next generation networks to efficiently use available spectrum in order to increase the bandwidth efficiency. Wireless Medical Networks helps monitoring of patient's vital signs using small sensors deployed on the body of patient. Cognitive radio can be effective in this network as well. 

\section{Related Work}
Time synchronization is required for coordination in every Cognitive Radio Network. Efficient implementation of CRN depends on time synchronization during phases of dynamic spectrum management. Furthermore, time synchronization also overcomes wireless channel issues such as shadow fading and interference \cite{b16, b17, b13, b14, b15}. Every CRN has its own requirements in terms of precision and accuracy. The minimum time error that a CR node can have with the master node defines the required precision and this time error must be minimized. Energy efficiency is the key factor required for any time synchronization protocol as it increase the lifetime of CR node. 

Time synchronization includes two types of synchronizations namely phase and frequency. Phase offset between two clocks is the difference of actual values between two clocks. Frequency offset known as skew is the difference between speeds of the clocks. Phase synchronization is usually done by using one timestamp and correction of phase, however skew correction requires multiple timestamps to estimate and correct the speed. The correction of clocks can be of three types. First approach is to instantaneously correct the phase of the clock which can cause gaps between time. Second approach is to adjust the speed of the clock to overcome the phase offset quickly. Our proposed scheme uses this approach. Third approach is to let the clock run untethered and a correction table is maintain which continuously compensates for offset. 

The detailed analysis of wireless network time sync protocols and Cognitive Radio Network time sync protocols are explained below. It also elaborates that why wireless network protocols are not helpful in synchronizing a cognitive radio network.

Many algorithms are proposed for wireless networks e.g, Reference Broadcast Synchronization (RBS), Timing-sync Protocol for Sensor Network (TPSN), Time-Diffusion Protocol (TDP) and Flooding Time Synchronization Protocol (FTSP) (\cite{b18, b19, b20, b21}). RBS tries to ensure that latency is minimized using direct  receiver to receiver synchronization. Each node sends reference broadcast beacons to its neighbors. These neighbors estimate the non deterministic values related to packet like transmit time, received time and time required for propagation. In this way, a table is generated which has clock values for all the network and then let the clocks run untethered. TPSN is a sender-receiver based synchronization protocol that creates tree of the nodes and synchronize them in two phases. A tree is created in first phase known as level discovery phase in which each node is assigned with a level and its master node. In second phase all nodes has to synchronize with only higher level nodes which is known as synchronization phase. In FTSP, a master node floods the time information to all the nodes. These receiving nodes simply correct their clock values by comparing the time information received. TPSN and FTSP is a promising protocol though it is not designed for CRN. These protocols are not sufficient because CRN requires channel handoffs and has adaptive features \cite{b14, b22}. Few protocols that are specifically designed for the adaptive nature of CRN are CR-Sync \cite{b15} and BSynC \cite{b14}.

CR-Sync is based on TPSN. It creates a tree structure and assign levels to each node within CRN. Synchronization is achieved by exchanging multiple timestamps between each parent and its children and eventually synchronizes the whole network \cite{b23}. In contrast with TPSN, CR-Sync is self-adaptive, fault tolerant and consider mobility aspects that is why it is suitable for CRN, however CR-Sync is not robust to root node failures.

Bio-inspired SynChronization (BSynC) proposed in \cite{b14} synchronizes pairs of nodes in real time. BSynC divides the nodes in CRN as Master Node, Ordinary Nodes, Reference Ordinary Nodes and Neighbor Nodes. Master Node is equipped with a master clock to which whole network synchronizes itself. Ordinary Node is the one which aims to get itself synchronized in the network. Reference Ordinary Node is an ordinary node nominated by another ordinary node to relate the values of their time and achieve synchronization together. Neighbor Node is a CR node inside a radio coverage of another CR node. After this division BSynC can be implemented by following two procedures i.e., Request to Synchronization Procedure (RSP) and Time Adjustment Procedure (TAP). In RSP, Master Nodes initialize the synchronization procedure by sending two types of messages to the Neighbor Nodes. These messages contains the available radio frequency channels, timestamp and Master Node identification number. In TAP, the receiving nodes create the Master Node as its Reference Ordinary Node. Each receiving node compare its clock value to the received timestamp and synchronizes in time by correcting clock. This method is followed though out the hierarchy and repeated number of times to synchronize in time phase and skew \cite{b14}.

The key contributions of the work is as follows: 
\begin{enumerate}
	\item  To the best of our knowledge, all of the current protocols for CRN requires multiple timestamp exchange to extract the time skew (\cite{b14, b15, b18, b19,b20,b21}). First time symbol timing recovery method of physical layer is used to extract the time skew.
	\item The method proposed in this paper extract the skew from just one timestamp with the help of symbol timing recovery  (\cite{b24, b25, b26, b27}).
	\item The proposed method is successfully applied to synchronize the CRN.
\end{enumerate}

\section{Proposed Methodology}

In CRN, each node maintains a register at the application layer which keeps the track of time. This application layer register is usually referred to as application layer clock, used for time synchronization within the network. The application layer register/clock keep on incrementing after counting the oscillations of quartz crystal embedded in hardware of each node in the CRN. Few of the current protocols are discussed in previous section that tries to correct this clock. However each protocol needs additional message circulation for time synchronization within the network. That is resource hungry and can consume a lot of energy. 

Similarly at the physical layer, whenever two nodes need to communicate, they need symbol timing recovery for the understanding of message. That timing recovery can be modelled as a clock sampling at optimum instant. That clock is based on same quartz crystal with in that node on which application layer clock is derived. Methodology proposed here utilizes symbol timing recovery information and corrects the application layer clock offset without using any additional message exchanges, causing the network to be more energy efficient.

\begin{figure}
	\centering
	\includegraphics[width=3.5in]{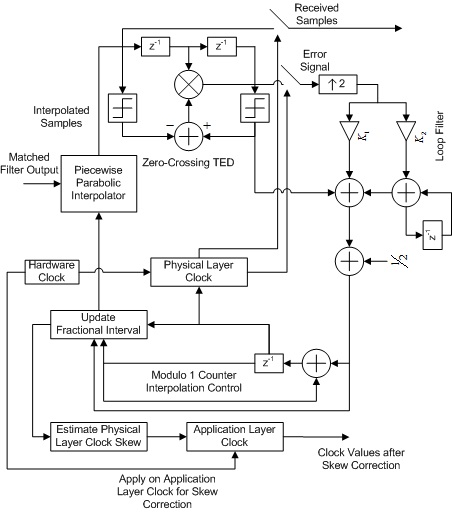}
	\caption{Cross layer timing synchronization system for Cognitive Radio Networks.}\label{TimeSyncSys}
\end{figure}

This proposed cross layer approach can be modeled as in Fig. \ref{TimeSyncSys}. At the receiving node, this model needs to be implemented using the following step by step procedure.\\\\
\textbf{Step-1:} The data from the match filter is interpolated by a piecewise parabolic interpolator.\\\\
\textbf{Step-2:} The interpolator is used to create a new sample at the desired instance which eventually causes sampling clock to synchronize. However, initially this instance is taken at random and latter updated by the interpolation control. 
 \\\\
\textbf{Step-3:} The interpolated samples are fed to Zero-Crossing Timing Error Detector which tries to locate the zero crossing of symbols and generates a corresponding error signal. That error signal is a measure of the difference left between current zero crossing of symbols and desired zero crossing. \\\\
\textbf{Step-4:} A loop filter further tries to track this error and instruct the modulation control to update the fractional interval on which next samples will be interpolated. The loop filter parameters $K_1$ and $K_2$ are adjusted such that it takes less tracking and acquisition time and have less variation in the output.
\\\\
\textbf{Step-5:} The interpolation control locates the offset that need to be adjusted in upcoming samples and also instruct that after which sample the new interpolated sample should be placed.\\\\
\textbf{Step-6:} Tracking the fractional interval can result in an estimate of physical layer clock skew. Skew estimation can be performed by any of the methods available in literature \cite{b25, b26, b27}. Clock skew estimated here using fractional interval and then sampling rate can be computed as,
\begin{equation}
f_{s}=\left ( 2+m \right )f_{d}, \label{eq_6}
\end{equation}
Where $f_s$ is adjusted sampling rate, $f_d$ is symbol rate and $m$ is the slope of the fractional interval.

Another way to estimate the skew is by Least Square (LS) method. A straight line needs to be mapped on the fractional interval values $y_i$ and error is minimized, which is given by,

\begin{equation}
e= \sum_{i=1}^{N}\left ( y_i-\widehat{y_i} \right )^{2}, \label{eq_7}
\end{equation}

Where $\widehat{y_i}$ represents the estimated line given by $y=mx+c$ and for $m$ and $c$ we have,

\begin{equation}
\centering
\begin{bmatrix}m\\c\end{bmatrix}=\left ( X^{T}X \right )^{-1}X^{T}Y, \label{eq_8}
\end{equation}

\textbf{Step-7:} This physical layer estimate of clock skew is now applied to application layer clock to correct the clock at the application layer. Hence there is no need of any additional time stamp to estimate the application layer clock skew.\\\\

\begin{figure}[H]
	\centering
	\includegraphics[width=2.5in]{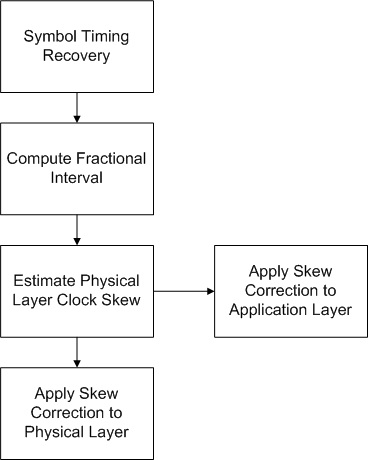}
	\caption{Cross Layer Time Synchronization.}\label{nmodel}
\end{figure}

A flow diagram is shown in Fig. \ref{nmodel} to further elaborate the proposed system discussed above.

The proposed method does not require any additional communication overhead because the sequence transmitted to extract the timing frequency offset is part of every communication protocol at physical layer. However an insignificant amount of computational overhead increases which is required to map the physical layer skew into application layer skew.

\section{Simulation Results}

For the proof of the proposed method, two synchronization systems are simulated. One synchronization system is for the application layer and one is for the physical layer (symbol timing recovery). Both systems estimate for clock skew of a node. It proves that proposed cross layer can be applied for the correction of application layer clock skew as skew comes out to be same for both physical and application layer.

Physical layer system is a simple binary PAM system that generates random symbols. Symbol rate is set to be $1000$ symbols/sec. Samples per symbols are set to be $8$. The transmitter simulates the up-sampling of PAM symbols and then pulse shaping using square root raise cosine with $50$ percent excess bandwidth. This data is assumed to be transmitted over a noise free channel. Symbol timing recovery system is applied at the receiver side, which utilizes ZCTED to compute timing error. The output of ZCTED is fed to piecewise parabolic interpolator which interpolates the samples and then fed to loop filter and interpolator control (the same components that are in proposed cross layer design). Symbol timing recovery system estimates the physical layer clock skew. 

To simulate application layer synchronization system, clocks are left to run at both transmitter and receiver side for the same duration ($500$ seconds). These clocks are modeled on the oscillations of the quartz crystal (sine wave). Counting specific number of oscillation increases a value to application clock. The frequency offsets can be seen in the oscillations produced. Then application layer clock skew is estimated using Least Square on multiple clock values.

\begin{figure}[H]
	\centering
	\includegraphics[width=3.5in]{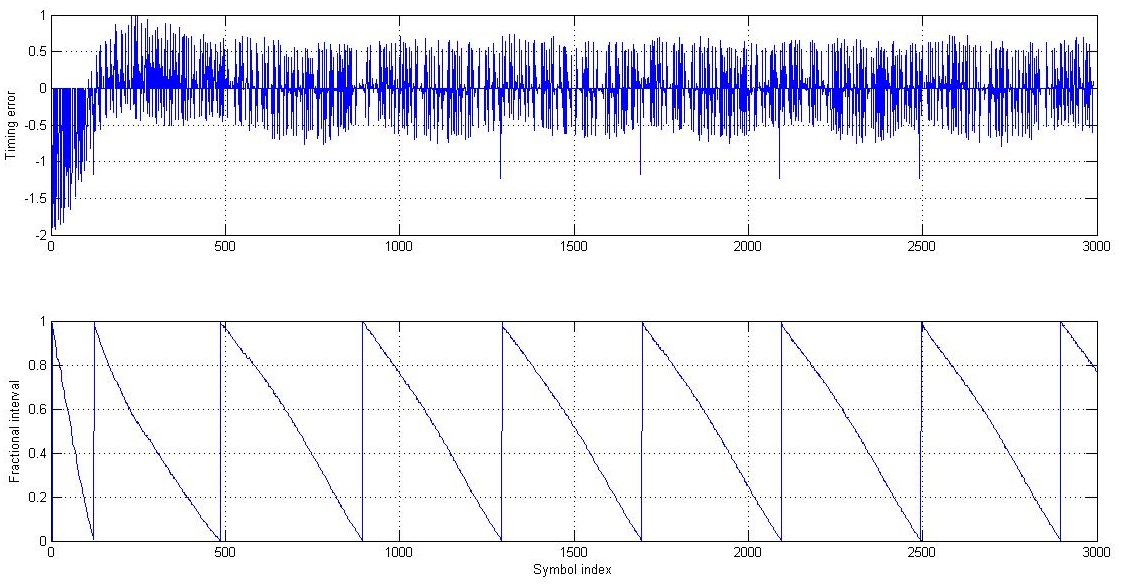}
	\caption{Timing error and fractional interval shows skew of $-1.2484 \times 10^{-3}$.}\label{nfig8}
\end{figure}

Simulation (-ve skew): 
Using the physical layer clock synchronization system discussed above, a -ve skew of $-1.2484 \times 10^{-3}$ is assumed. The simulation results are shown in Fig. \ref{nfig8} which shows estimated skew to be $-1.2438 \times 10^{-3}$.

For the application layer time synchronization, the same skew is assumed as transmitter and receiver are the same nodes. Transmitter application layer clock counts the oscillation till $100000$ and receiver application layer clock counts till $100124$ for the same duration i.e. $500$ seconds, while exchanging there clocks. The skew estimated using LS is $1-((100000-1)/(100124-1)) = -1.2385 \times 10^{-3}$ that comes out to be approximately equal to clock skew of physical layer. The result shows that the physical layer clock skew  ($-1.2438 \times 10^{-3}$) and application layer clock skew ($-1.2385 \times 10^{-3}$) are approximately identical. Now if we apply cross layer approach then physical layer skew estimate can correct the application layer clock skew.

\begin{figure}[H]
	\centering
	\includegraphics[width=3.5in]{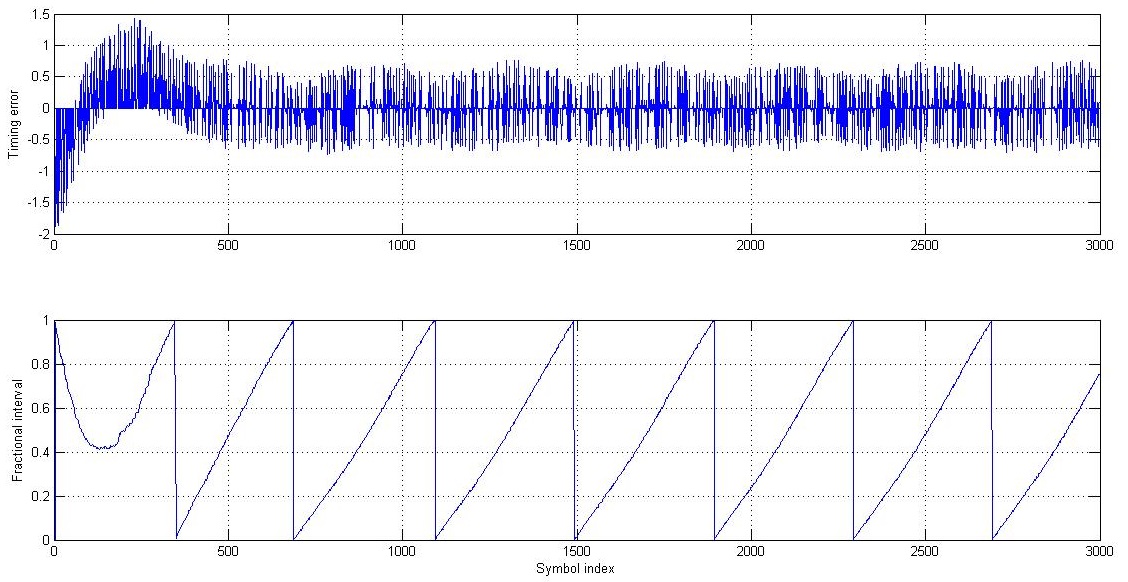}
	\caption{Timing error and fractional interval shows skew of $1.2469 \times 10^{-3}$.}\label{nfig7}
\end{figure}

Simulation (+ve skew): 
Assumption of clock skew present is $1.2500 \times 10^{-3}$ between the transmitter and receiver node. The physical layer simulation results are shown in Fig. \ref{nfig7} which shows estimated skew to be $1.2469 \times 10^{-3}$.

For the application layer time synchronization, the same skew is assumed as transmitter and receiver are the same nodes. Transmitter application layer clock counts till $100000$ and receiver application layer clock counts till $99875$ for the same duration i.e. $500$ seconds, while exchanging there clocks. The skew estimated using LS is $1-((100000-1)/(99875-1)) = 1.2516 \times 10^{-3}$ that comes out be  approximately identical to clock skew of physical layer. The result shows that physical layer clock skew ($1.2469 \times 10^{-3}$) and application layer clock skew ($1.2516 \times 10^{-3}$) are approximately same. Now if we apply cross layer approach then physical layer skew estimate can correct the application layer clock skew.

\section{Conclusion}

Method proposed in this paper relies on the fact that the physical layer clock ticks and application layer clock ticks are eventually operated from quartz crystal oscillations of the node and hence have a same skew offset. It can also be seen in the simulation results that both clock skews are same. Hence, skew estimate of physical layer clock can be directly applied for application layer clock correction of the nodes participating in communication within a cognitive radio network. This eliminates the need of any application layer protocol to be used in cognitive radio network and can reduce the energy requirements to a greater extent or otherwise these application layer protocols can be implemented with the proposed method in order to achieve better efficiency.

\section*{Acknowledgments}
This work was supported by the National Research Foundation of Korea(NRF) grant funded by the Korea government(MSIP) (No. 2017R1A2B4001801).

%%%%%%%%%%%%%%%%%%%%%%%%%%%%%%%%%%%%%%%%%%%%%%%%%%%%%%%%%%%%%%%%%%%%%%%%%%%%%%%%%%%%%

%\bibliography{ijmso}
%bibliographystyle{unsrt}
%\bibliographystyle{alpha}

%\newpage

\end{document}